\documentstyle[aps,psfig,preprint]{revtex}

\begin{document}
\title
{\bf Chemical equilibration and thermal  dilepton production from
the quark gluon plasma at finite baryon density}
\author
{D. Dutta, K. Kumar, A. K. Mohanty and R. K. Choudhury}
\address{ Nuclear Physics Division, Bhabha Atomic Research Centre,
Bombay-400085}
\date{\today}
\maketitle
\begin{abstract}
The chemical equilibration of a highly unsaturated
quark-gluon plasma has been studied at finite baryon density.
It is found that in the presence of small amount of baryon density,
the chemical equilibration for gluon becomes slower and the temperature
decreases less steeply as compared to the baryon free plasma.
As a result, the space time integrated yield of dilepton
is enhanced if the initial temperature of the plasma is held fixed.
Even at a fixed initial energy density, the suppression of the dilepton yields
at higher baryo-chemical
potential is compensated, to a large extent, by the slow cooling of the plasma.
\end{abstract}
\section{Introduction}
One of the important objectives of all the future collider experiments
at RHIC and LHC is to detect the new state of matter called quark gluon
plasma (QGP) which is expected to be produced
during the ultra-relativistic
heavy ion collisions. Such an exotic state, if formed, will cool during
expansion till it  reaches a critical temperature $T_c$ where the QGP
will be transformed to the hadron phase via
a first or second-order phase transition. The hadrons will then cool down further
to the freeze-out temperature  $T_f$ where they cease to interact with each other
and fly away to the detectors. During the process of the thermal expansion,
 photons and dileptons
are  produced directly from the plasma as well as from the hadron phases.
These thermal photons and dileptons are considered to be
 ideal probes  for the detection and study of subsequent
evolution of the QGP \cite{1}
 as they leave the  plasma without any  interaction.
\paragraph*{}
In the standard scenario \cite{2}, the quark gluon plasma  formed
during the  collision is expected to thermalize
 in a typical time scale of 1 fm/c.
However, some recent works \cite{3,4} suggest that due to high initial parton density
(mostly gluonic) at RHIC and LHC energies, the plasma
may attain kinetic equilibrium in a very short
time ($\tau \approx 0.3 fm - 0.7 fm$)  but it may be far  from chemical
equilibrium. Since the initial parton plasma is mostly gluonic,
many more quarks and anti-quarks are  needed in order
to achieve chemical  equilibrium.
Earlier studies \cite{5,6,7} have shown that  a chemically
non-equilibrated plasma cools faster as compared to equilibrated plasma,
which follows Bjorken's scaling law ($ T^3 \tau$=const ). In the case of
non-equilibrated plasma, additional energy is consumed in
producing more quarks and anti-quarks in approaching chemical equilibrium due to
which the cooling is accelerated as compared to the Bjorken's scaling law.
Since the plasma cools faster, it may not attain chemical equilibrium
before the temperature reaches the critical value $T_c$.
Even if one includes transverse expansion, the large velocity gradient
may drive the system further away from the chemical equilibrium as shown by
\cite{8}.

The effect of chemical equilibration
 on thermal photon and dilepton emission has been studied
by several authors  \cite{7,8,9,10,11,12,13}. In these studies, it was assumed that the
nucleus- nucleus collision is fully transparent, leading to the formation
of a baryon-free plasma. However, some of the recent works have suggested
that even at RHIC energies the colliding nuclei may not be fully
transparent and some amount of baryon stopping may occur
particularly at higher rapidities . For $Au+Au$ collisions at 200 A GeV,
the baryo-chemical potential $\mu_B$ may be of the order of T
 for $y\approx 2$ \cite{14}, which
is the region of interest of the PHENIX detector for di-muon measurements \cite{15}.
The presence of finite baryon density may affect the process of chemical
equilibration and the rate of cooling. This will, in turn,
affect the thermal dilepton yields, which are the potential probes
for the detection of the expanding quark gluon plasma.

The present work is aimed at studying these effects, by considering
the dynamical evolution of a baryon rich plasma undergoing chemical equilibration.
It is seen that the rate of cooling of the plasma
slows down  in the presence of $\mu_B$ compared to a baryon-free plasma.
 The calculations also show that
the rate of dilepton production is suppressed at finite baryon density, but
the time integrated thermal yields are rather unaffected
due to slower cooling of the plasma in the presence of small amount of
the baryon density.

\section{Chemical equilibration}
We assume a quark gluon plasma which has thermally equilibrated
but is far off from chemical equilibrium.
The distribution functions for quarks (anti-quarks) and gluons for such an
unsaturated plasma can be described by Juttner  distributions given by
\equation f_ q(_{\bar q})  = \frac{\lambda_ q(_{\bar q})}
{\lambda_ q(_{\bar q}) + e^{(p \mp\mu)/T}}=
\frac{\lambda_{q({\bar q})}  e^{\pm x}}
{\lambda_{q({\bar q})} e^{\pm x}+ e^{p /T}} ~~~~
;~~~~~f_ g = \frac{\lambda_ g}{e^{p/T}-\lambda_g}
\endequation
where $x=\mu/T$. The fugacity factor $\lambda_i $ ($i=q,\bar{q}$ and g) gives the measure of
the deviation of the distribution functions from the equilibrium values
 and $\mu$ (= $\mu_B$/3) is the quark-chemical potential.
The chemical equilibrium is said to be achieved when $\lambda_i \rightarrow 1$.
It may be mentioned here that one
can also define the quark and anti-quark distribution functions
using  a different definition for fugacities $\lambda_Q$ and $\lambda_{\bar Q}$
given by
\equation \lambda_{Q({\bar Q})}=e^{\pm x}\lambda_{q({\bar q})} \endequation
In that case, one does not use the quark chemical potential $\mu$ explicitly.
However,
the definition (1) is quite convenient, since  at
equilibrium ($\lambda_q$=$\lambda_{\bar q}$=1), the chemical potential associated with $\lambda_i$ vanishes
but the baryo chemical potential still exists. Further,
we have taken $\lambda_{q} = \lambda_{\bar q}$ at all values of $\tau$ so that
when $\mu\rightarrow  0$, $\lambda_{Q} = \lambda_{\bar Q} =\lambda_q$ resulting in a baryon
symmetric matter.

\subsection {Rate equations}
In general, chemical reactions among partons can be quite complicated because
of the possibility of initial and final state gluon radiations. However,
we restrict our considerations to the following dominant
reaction mechanisms \cite{5,7} for the equilibration of parton flavours,\\

~~~~~~~~~~~~~~~~~~~~~~~~~~~~~~~$gg \rightleftharpoons ggg$~~~~~~$gg \rightleftharpoons q \bar q$\\
The evolution of the parton densities are governed by the master equations
\equation \partial_\mu (n_gu^\mu) = (R_{2\rightarrow 3} -R_{3\rightarrow 2})
-(R_{g\rightarrow q} - R_{q\rightarrow g}) \endequation
\equation \partial_\mu (n_qu^\mu) = \partial_\mu (n_{\bar q} u^\mu)=
(R_{g\rightarrow q} - R_{q\rightarrow g})\endequation
where $R_{2\rightarrow 3}$ and $R_{3\rightarrow 2}$ denote the rates for the
process
$gg \rightarrow ggg$ and its reverse and $R_{g \rightarrow q}$ and $R_{q \rightarrow g}$
for the process $gg \rightarrow q \bar q$ and its reverse respectively.
In case of a baryon rich plasma, the presence of quark-chemical potential
 $\mu$, affects the process $gg \rightleftharpoons q \bar q$ directly. We,
 therefore, study the $\mu$ dependence of this process more explicitly.
The  rate equations for this process are given by\cite{16}
$$ R_{g\rightarrow q}= \frac{1}{2} \int \frac{d^3p_1}{(2\pi)^3 2 E_1}
\int \frac{d^3p_2}{(2\pi)^3 2 E_2} \int \frac{d^3p_3}{(2\pi)^3 2 E_3}
\int \frac{d^3p_4}{(2\pi)^3 2 E_4} (2\pi)^4 \delta^4(p_1+p_2-p_3-p_4) ~\times
$$
\equation
\sum |M_{gg\rightarrow q\bar q}|^2 f_g(p_1)f_g(p_2)[1-f_q(p_3)][1-f_{\bar q}(p_4)]
\endequation
and
$$ R_{q \rightarrow g}= \frac{1}{2} \int \frac{d^3p_1}{(2\pi)^3 2 E_1}
\int \frac{d^3p_2}{(2\pi)^3 2 E_2} \int \frac{d^3p_3}{(2\pi)^3 2 E_3}
\int \frac{d^3p_4}{(2\pi)^3 2 E_4} (2\pi)^4 \delta^4(p_1+p_2-p_3-p_4)~\times
$$
\equation
\sum |M_{q \bar q\rightarrow gg}|^2 f_q(p_3)f_{\bar q}(p_4)[1+f_g(p_1)][1+f_g(p_2)]
\endequation
In (5), the squared matrix element, summed over spin and color,
$\sum |M|^2$ is weighted by two
gluon distribution functions $f_g$ for the initial states. The factor $
[1-f_q][1-f_{\bar q}]$ indicates Pauli blocking for the final states. In the reverse process (6),
the rate is weighted by the distribution functions of quarks
and anti-quarks
for the initial states and the gluon final states each gain an enhancement
factor $[1+f_g]$ due to Bose-Einstein statistics. The factor of 1/2 accounts
for the identity of the two gluons.

 Using the identity
 \begin{equation}
 [1-f_q][1-f_{\bar q}]=
 \frac {f_q(p_3) f_{\bar q}(p_4)}
 {\lambda_q \lambda_{\bar q}}
 e^{(p_3+p_4)/T} \end{equation}
 and
 \begin{equation}[1+f_g(p_1)][1+f_g(p_2)]=
 \frac {f_g(p_1) f_g(p_2)}{\lambda_g^2}
 e^{(p_1+p_2)/T} \end{equation}

 it may be shown that the $\mu$ dependence in the rate equations
 (5) and (6) for the process $gg\rightleftharpoons
q\bar{q}$ basically comes through the product of quark and anti-quark
distribution functions $f_q f_{\bar q}$.

Fig. 1(a) and (b) show the variation of the product $f_q f_{\bar q}$
with momenta p and  $\lambda_q$ respectively for different values of $x$.
Fig. 1(a) shows the product $f_q f_{\bar q}$ as a
function of p at a few typical values of $x~(=\mu/T)$ for
$\lambda_q=$ 0.1 and for temperatures T=0.57 GeV and T=0.2 GeV.
Fig. 1(b) shows the $f_qf_{\bar q}$ as a function of $\lambda_q$
at temperature T= 0.57 GeV and p=0. It is seen (Fig 1a) that
as the momentum increases or the temperature decreases, the $e^{p/T}$ factor
in the denominator of (1) dominates and the product $f_q f_{\bar q}$
becomes less sensitive to $\mu$.
Here, we have used the same values of p  for both quark and anti-quark
distribution functions, although their momenta $p_q$ and $p_{\bar q}$ can be
quite different.
However, for any other combinations of $p_q$ and $p_{\bar q}$, the deviation
will not be more than that observed for p=0.
 We have taken $p_q=p_{\bar q}=0$ in Fig. 1(b)
to show the maximum amount of deviation of this product as a function of
$\lambda_q$. It can be seen from Fig 1(b), that the above product is not
very sensitive quark chemical potential
if the plasma is highly unsaturated ($\lambda_q$  $\le$ 0.1).
This feature has an important relevance for the collisions at RHIC
energy as the chemical equilibration is not achieved by the time
$T$ drops to $T_c$ and the quark fugacity remains much below unity
($\lambda_q \approx $0.1) \cite{5}.
The nearly $x$ independence of the above product is also evident from
eq. (1).
For small values of $\lambda_q$, the contribution from the factor $\lambda_q e^{\pm x}$
in the denominator is not very significant if $x$ is small. The $x$
dependence of the quark and anti-quark distribution functions mainly arises
due to the $e^{\pm x}$ factor in the numerator. Since these exponential
factors get cancelled in the product, $f_q f_{\bar q}$ will have weak
$x$ dependence at small baryon density if the plasma is highly unsaturated.
So the integrals (5) and (6)
will not depend explicitly on $\mu$  except through the thermal
quark mass $m_q^2$ which is generally used as a cut-off parameter in the
evaluation of the integral to avoid the divergence.
The thermal quark mass appropriate for a non-equilibrium plasma
is given by \cite{17,18}
\equation
m_q^2 = \frac{~g^2}{3\pi^2} \int_0^\infty dp~p~[2~f_g+f_q+f_{\bar q}]
\endequation
Since the gluon distribution function has no $\mu$ dependence, the rate for
the process $gg \rightleftharpoons ggg$ will depend on $\mu$ through the Debye
screening mass $m_D^2$ which is
used to avoid the divergence in both the scattering cross sections and
the radiation amplitude and can be given by \cite{19}
\equation
m_D^2 = \frac{3~g^2}{\pi^2} \int_0^\infty dp~p~[2~f_g+N_f(f_q+f_{\bar q})]
\endequation

In the above, $N_f$=2.5 is the dynamical quark flavour and $g^2 = 4\pi\alpha_s$ is the
strong coupling constants.
The above rate equations have been solved by using the distribution function
as discussed below.

\subsection{Distribution function}

In order to evaluate  energy, number densities, thermal quark mass and Debye screening mass
we consider
the approximations for the quark and anti-quark distribution functions given by
\equation
f_ {q({\bar q})} =
\frac{\lambda_{q({\bar q})}  e^{\pm\frac{\mu}{T}}}{\lambda_{q({\bar q})} e^{\pm \mu/T}+ e^{p /T}}
\approx \frac{\lambda_ {q({\bar q})} e^{\pm x}}{1+ e^{p /T}}= \lambda_{q({\bar q})} e^{\pm x} f^{eq}_{q (\bar{q})}
\endequation
and  gluon distribution function as
\equation  f_g=\lambda_g f_g^{eq} \endequation
where $f_{q({\bar q})}^{eq} = (1+e^{p/T})^{-1} $ and $f_g^{eq} = (e^{p/T}-1)^{-1} $.
For subsequent reference, we will call it as Modified Fermi Dirac type (MFD) approximation.
It should be mentioned here that the  most commonly used approximation
(extended to non-zero $\mu$) is given by
\equation
f_ {q({\bar q})} = \frac{\lambda_ {q({\bar q})}}{1 + e^{(p \mp \mu)/T}}
\endequation
The above approximation  becomes Fermi Dirac  distribution in the limit
$\lambda_q \rightarrow  1$ and we will refer it as FD type approximation.
It has been shown in appendix A that this FD approximation  is not strictly applicable
in case of an unsaturated plasma (i.e. for $\lambda_i< 1$), though it has been used in all the
earlier works \cite{5,6,7,8,13} for baryon free plasma
(i.e. $\mu=0$).  The MFD approximation (11) becomes FD type when $\mu \rightarrow 0$.
We would also like to compare the MFD approximation with the
Boltzmann (BM) approximation given by
\equation f_ q(_{\bar q})  = \lambda_ q(_{\bar q})e^{-(p \mp\mu)/T}\endequation
The BM approximation is found to be closer to the anti-quark distribution
function at finite baryon density. Comparison between different approximations
has been discussed in appendix A.

We have evaluated the thermal quark mass $m_q^2$ numerically using (9)
for various approximations as well as for the Juttner distribution.
Fig. 2  shows $m_q^2/T^2$ as a function of $\lambda_q$ at two
different values of $x$.
Here, the gluon contribution to thermal mass has not been included
since we are interested only in  examining the difference arising from
the use of different approximation for the quark distribution function.
It may be seen that the FD approximation
coincides with the Juttner distribution only in the limit $\lambda_q \rightarrow 1$. However,
in the region of interest i.e. $\lambda_q $ up to 0.1 or 0.2, the deviation
is quite significant. Similarly, the Boltzmann approximation  agrees with the
Juttner distribution only at very small values of $\lambda_q$ while MFD approximation
agrees  with the Juttner distribution up to $\lambda_q \approx 0.35$ even
at $x=1.5$. Since the plasma
is highly unsaturated to begin with and the quark fugacity remains much below unity,
we will  use the MFD approximation  in the subsequent calculations.
The advantage of using this approximation is that we can  retain the
factorisation  used in \cite{5} for the RHS of the rate equations (3) and (4)
with the replacement of $\lambda_{q(\bar q)}$ with $\lambda_{Q(\bar Q)}$
(see appendix B for detail). The rate equations are then solved in the following way.

\subsection{Formalism}

We assume the system to undergo a purely boost
invariant expansion. Using the MFD approximation, (3) and (4) can be written as

\begin{equation} \partial_\mu (n_g u^\mu)= \frac {\partial n_g}{ \partial \tau}~+~\frac { n_g}{ \tau}\\
=n_g R_3(1-\lambda_g)-2 n_g R_2 (1- \frac {\lambda_Q \lambda_{\bar Q}}{
\lambda_g^2}) \end{equation}

\begin{equation}\partial_\mu (n_q u^\mu)=\frac {\partial n_q}{\partial \tau}~+~\frac {n_q}{\tau}\\
=n_g R_2 (1-\frac {\lambda_Q \lambda_{\bar Q}}{\lambda_g^2}) \end{equation}

where $R_2=\frac{1}{2}\sigma_2 n_g$ and $R_3=\frac{1}{2} \sigma_3 n_g$ are the
density weighted cross sections
for the process $gg \rightleftharpoons qq$ and  $gg \rightleftharpoons ggg$
respectively.
Similarly, the equation for the conservation of energy and momentum can be
written as
\begin{equation}   \frac {\partial \epsilon }{ \partial \tau} + \frac {\epsilon+p} {\tau }=0 \end{equation}
where the viscosity effect has been neglected \cite{20} .
In case of an ideal fluid, 3p =$\epsilon$.

Using MFD approximation for quark, anti-quark (11) and gluon distribution functions (12),
the energy density can  be written as:

\begin{equation}   \epsilon=T^4[a_2\lambda_g+b_2(\lambda_q e^x+\lambda_{\bar q} e^{-x})] \end{equation}

   with $a_2=8 \pi^2/15$;~~ $b_2=N_f(7 \pi^2/40)$~; where $N_f$=2.5 is the
   dynamical quark flavours. Similarly, the number densities
   for gluon, quark and anti-quark are:

\begin{equation}  n_g=\lambda_g a_1 T^3 ;~~~ a_1=\frac{16}{\pi^2}~~ \zeta (3) \end{equation}
\begin{equation}  n_q=\lambda_q b_1 T^3 e^x ;~~ b_1=\frac{9}{2\pi^2}~~ \zeta (3) N_f \end{equation}
\begin{equation}   n_{\bar q}=\lambda_{\bar q} b_1 T^3 e^{-x} \end{equation}
From the conservation of baryon number, one gets $\partial_\mu(n_Bu^\mu) =0$
which results in
\equation n_B\tau = (n_q-n_{\bar q})\tau = const \endequation

Using the expressions for the number and energy densities, the  eq. (15,16) and
(17,22) can now be written as

\begin{equation} \frac{{\dot \lambda}_g}{ \lambda_g }+ \frac{ {3 \dot T}}{T} + \frac{1}{\tau} =
         R_3(1-\lambda_g)-2 R_2 ( 1- \frac{\lambda_Q  \lambda_{\bar Q} }
         { \lambda_g^2}) \end{equation}

\begin{equation} \frac{{\dot \lambda}_Q}{ \lambda_Q }+ \frac{ {3 \dot T}}{T} + \frac{1}{\tau} =
         R_2 \frac{ a_1}{b_1} \frac{\lambda_g}{ \lambda_Q} ( 1- \frac{\lambda_Q  \lambda_{\bar Q} }
         { \lambda_g^2}) \end{equation}

\begin{equation} {\dot \lambda}_{\bar Q} ={\dot { \lambda}_Q }+
                   \frac { \lambda_Q -  \lambda_{\bar Q }}{\tau}    +
         \frac{ 3 {\dot T}}{T} (\lambda_Q -  \lambda_{ \bar Q})
 \end{equation}

 \begin{equation} \frac {3 {\dot T}} {T}+ \frac{1}{\tau} = -~~\frac{3}{4~A_t}
 [  a_2 {\dot \lambda}_g
 +  b_2 {\dot \lambda}_Q
 +  b_2 {\dot \lambda}_{\bar Q}   ]
 \end{equation}

where $$ A_t =  a_2 \lambda_g + b_2 (\lambda_Q + \lambda_{\bar Q}) $$

The above equations are solved numerically estimating $R_2$ and $R_3$ in a similar
way as that of ref. \cite{5,7}.
Following ref. \cite{5}, the  rate $R_2$ is written as
\begin{equation}
R_2 \approx .24 N_f~\alpha_s^2~ \lambda_g ~T ~ln(\frac{1.65}{\alpha_s \lambda})
\end{equation}
where $\alpha_s(=0.3)$  is the strong coupling constant.
The factor $\lambda= (\lambda_g+\cosh x\lambda_q/2)$
arises due to the thermal quark mass $m_q^2$ given by
\begin{equation}
 m_q^2=\frac{4\pi\alpha_sT^2}{9}[\lambda_g+\lambda_q \cosh x/2]
\end{equation}
which has been evaluated
using  MFD  distribution functions.
Following ref. \cite{7}, we  estimate $R_3$ numerically

\equation
R_3/T = \frac{32}{3 a_1} \frac{\alpha_s}{ \lambda_g}\left[\lambda_g + \lambda_q\frac{N_f}{6} ~ \cosh x\right]^2
\left[1+\frac{2}{9}\frac{m_D^2}{T^2}\right]^2
I(\lambda_g,\lambda_q,x)
\endequation
where $I(\lambda_g,\lambda_q,x)$ is a function of $\lambda_g,\lambda_q$ and $x$,
\begin{eqnarray}
I(\lambda_g,\lambda_q,x)& = &\int_1^{\sqrt s \lambda_f} dx \int_0^{s/4 m_D^2}
dz \frac{z}{(1 + z)^2}\nonumber\\
&&\left[ \frac{\cosh^{-1}(\sqrt {x})}
{ x \sqrt{[x + (1+z)x_D]^2-4 x z x_D}}
 + \frac{1}{s \lambda_f^2} \frac{\cosh^{-1}(\sqrt x)}
 {\sqrt {[1+x (1+z)y_D]^2-4 x z y_D}} \right]
\end{eqnarray}
with $x_D=m_D^2~\lambda_f$ and  $y_D=m_D^2/s$, $s=18~T^2$
is the square of the average center of mass energy. The mean free path for elastic
scattering is given by $\lambda_f$
\equation
 \lambda_f^{-1} = \frac{9}{8}~ \alpha_s~a_1~ T~\lambda_g~
\left[\lambda_g +\lambda_q \frac{N_f}{6} ~ \cosh x\right]^{-1}\left[1+\frac{2}{9}\frac{m_D^2}{T^2}\right]^{-1}
\endequation
The above equations (29-31) are similar as that of used in ref. \cite{7}
except it has been rederived to have $x$ dependence through the
Debye screening mass $m_D^2$ given by
\equation
 m_D^2 = 4 \pi \alpha_S T^2 \left[\lambda_g +\lambda_q \frac{N_f}{6} ~ \cosh x\right ]
\endequation
derived using MFD distribution functions.
It may be noted here that even in case of a baryon free plasma ($x$=0), the
expression for $m_D^2$ still differs from $4\pi\alpha_s T^2 \lambda_g$
which has been used in ref. \cite{5,7}.
 The rates $R_2 $ and $R_3$ were calculated using the above equations.
Fig. 3 shows some typical results of the calculations  of
 $R_2/T$ and $R_3/T$   as a function of $\lambda_g$ for
 $x$=0.0, 1.0 and 1.5. The values of $\lambda_q=\lambda_{\bar q}$ were
fixed at $\lambda_g$/5 . As can be seen, the
equilibration rate $R_3$ initially increases and later on decreases
with increasing baryon density
whereas the rate $R_2$ is not affected much except at higher fugacities.
This can be understood from the eq. (27) where $R_2$ has a logarithmic
$x$ dependence whereas $R_3$ depends on $x$ more directly  through the Debye
screening mass $m_D^2$ (eq. 32).
Also shown in the same figure are the results for $R_3/T$ used in ref.
\cite{7} for x=0 (dot-dashed curve) with $m_D^2=4\pi \alpha_s T^2 \lambda_g$
which does not include the quark and anti-quark contributions.
It is seen that, with the inclusion of quark and anti-quark contribution in the
Debye screening mass, the rates $R_3/T$ are lowered for $\lambda_g > 0.2$, but
are enhanced for smaller $\lambda_g$ values .

\subsection{Results}
   The time dependence of $\lambda_g$, $\lambda_Q$, $\lambda_{\bar Q}$ and $T$
were obtained by solving the set of rate eqs (23-26) numerically by fourth
order Runga-Kutta method. We take the initial conditions from the
Hijing calculations, $T_0$= 0.57 GeV, $\lambda_{g0}$=0.09, $\lambda_{q0}$ =0.02
at $\tau_0$ =0.31 fm as used in \cite{5} and treat $x_0$ as a parameter.
We have carried out a parametric study to see the effect of chemical potential $x$ on
the chemical equilibration.

\paragraph *{}
Figs. 4(a-d) show the
temperature T, chemical potential $x=\mu/T$
and the fugacities ($\lambda_g, \lambda_q= \lambda_{\bar q}$)
as a function of $\tau$  at few typical values of $x_0$.
Whether the plasma is baryon free or baryon rich, a common feature is that
the process of chemical equilibration needs additional amount of energy
which makes the plasma cool more rapidly than predicted by Bjorken's
scaling $T^3\tau$=const (dotted line, Fig. 4a). However,
  the plasma cools less rapidly in the presence of chemical
potential  compared to  the baryon free plasma.
 Further, it is seen that the fugacity factors for gluon, quark and
anti-quark do not reach the equilibrium values by the time the temperature T
drops to  the critical value $T_c$ $\approx$ 0.2 GeV (Fig. 4(c) and (d)).
It is important to note that
the gluon equilibration rate slows down in the presence of finite
baryon density. On the other hand, the quark and anti-quark equilibration rates
increase slightly in the presence of finite baryon density.
As shown in Fig. 4(b), the chemical potential $x=\mu/T$ also
decreases with  $\tau$. The overall effect, however, is that the plasma cools
less rapidly in the presence of chemical potential as compared to the baryon free
plasma.

The different behaviour of $\lambda_g$ and $\lambda_q$ observed in
Figs 4(c) and 4(d) as a function of $\tau$ may be understood in the following
way. The $x$ dependence in the chemical equilibration arises due to the factors
$R_2$, $R_3$ and baryon density $(n_B)$
present in the rate equations (eq. 15 and eq. 16). As shown in Fig. 3, the rates $R_2$ and $R_3$ do not
depend on $x$ significantly.
It was found that even if $R_2$ and $R_3$ are made independent of $x$,
i.e. if we drop the term containing
$\lambda_q \cosh x $ in $m_q^2$ and $m_D^2$ completely,
the rate of chemical equilibration is not affected much.
Therefore, the slowing
down of the gluon equilibration rate is not due to $R_2$ and $R_3$, but due to
the dynamical evolution of the plasma which is affected in the presence of
baryon density (See Appendix C for detailed discussion).
The finite baryon density makes the cooling rate slower as the
plasma evolves with an additional constraint of baryon number conservation.
This also effects the plasma density which is mostly gluonic.
It may be mentioned here that even though $R_2$ and $R_3$
do not depend strongly on $x$, the rate of chemical equilibration
ultimately depends on how fast
the net parton production approaches zero, i.e. how fast the RHS of eq. (15) or eq. (16)
becomes zero. Since initially $\lambda_g > \lambda_q$ and $R_3 > R_2$, a slight
increase in gluon density produces more ggg as compared to $q \bar q$ or gg pairs
( i.e. more gain than loss in the gluon numbers as seen from eq. 15).
As a result, the gluon equilibration rate decreases where as the quark or anti-quark
equilibration rate practically remains unchanged until at a later time when it goes up
slightly. The decrease in gluon equilibration rate will consume less energy and
will make the plasma cool still slower.
Therefore, the
over all slowing down of the plasma cooling rate is due to both slower gluon
equilibration as well as a slower expansion rate of the plasma that results in
presence of baryon density.

We have also investigated the rate of cooling by increasing the baryon density
further, although our approximation breaks down at high $x$ values.
We find that the cooling rate never exceeds the Bjorken's limit which is
consistent with the fact that for an ideal plasma Bjorken's scaling is the
upper limit even in case of a baryon rich plasma.
This may  be verified directly from the conditions $s\tau $= const and
$n_B\tau$=const (s and $n_B$ are entropy and baryon density respectively)
without solving any rate equations. For an equilibrated plasma,
$x$ remains independent of $\tau$
and the
temperature follows the rule $T^3\tau$ = const.

It is known that the rate
of parton equilibration
is enhanced considerably due to various factors like higher
order gluon multiplication \cite{21,22}, temperature dependent coupling constants \cite{23} and
viscosity.
All these processes
need to be included for a complete understanding of the chemical equilibration
during the hydrodynamical evolution of the plasma. We do not include all the above
effects due to various complexities involved in the calculations. For example,
the perturbative calculations may not be valid if $\alpha_s$ is allowed to vary
with $\tau$ which may have large value at the end of the evolution \cite{23}. Similarly,
to calculate the rate for $gg \rightarrow (m-2)g$; $m>5$, one needs to understand the
complete space time structure of the multi gluon processes \cite{22}. Similarly, inclusion of dissipative
effects complicates the problem further.
Therefore, in order to isolate the effect of baryo-chemical
potential on the chemical equilibration
, we have considered in this work a non-viscous plasma which expands isentropically.
We do not include
any higher order gluon multiplication processes and also we use
a constant value for $\alpha_s$=0.3. The results on the dilepton production
yields are discussed in the following section.
\section {Dilepton production}
Thermal photons and dileptons are the ideal probes to test the
hydrodynamical evolution of a plasma created in heavy ion collisions at
ultra-relativistic energies.
It was shown in ref. \cite{14,17} that the thermal photon and dilepton rates are suppressed
at finite baryon density when the rates are calculated at a fixed energy density.
However, this suppression is primarily due to lower
initial temperature that results at higher baryon density.
We will show in the following
that the presence of finite baryon density
enhances the space-time
integrated thermal yield of dileptons, in spite of the fact that
the rate of production is suppressed \cite{14} .
We have considered here the dilepton production, however this study
can be extended to the thermal photons as well.

The dileptons are produced predominantly via the reaction $q^+q^- \rightarrow l^+l^-$.
We ignore the annihilation and the Compton like reactions in the present calculations
since their contributions may not be important for invariant masses above 1 GeV.
The dilepton production rate $dN/(d^4xd^4p)=dR/d^4p$ (i.e. the number of
dileptons produced per space time volume and four dimensional momentum volume)
is given by:
\equation \frac{dR}{d^4p} = \int \frac {d^3p_1}{(2\pi)^3}
\int \frac {d^3p_2}{(2\pi)^3} f_q(p_1) f_{\bar q}(p_2) v_{q \bar q}
\sigma_{q \bar q} ^{l^+l^-} \delta^4(p-p_1-p_2) \endequation
where
$v_{q \bar q}$ is the relative velocity between quark and anti-quark and
$\sigma_{q \bar q} ^{l^+l^-}$ is the total cross section for the reaction
$q \bar q \rightarrow l^+l^-$
\equation \sigma_{q \bar q} ^{l^+l^-} = \frac {4}{3} \frac {\pi \alpha^2}
{M^2} \endequation
$M=p^\mu p_\mu$ is the invariant mass of the dileptons.
The above  integral can be simplified to \cite{10}
\equation \frac{dR}{dM^2} = \frac{5}{24\pi^4}M^2\sigma(M^2)
\int_0^\infty dp_1 f_q(p_1) \int_{M^2/4p_1}^\infty
dp_2 f_{\bar q}(p_2) \endequation
We evaluate the above integral numerically using the exact
definition of quark and anti-quark distribution functions as given by (1).
Fig. 5(a) shows the plot of $dR/dM^2$ as a function of dilepton invariant mass
at fixed initial temperature T=0.57 GeV corresponding to an equilibrated ($\lambda_q=1.0$)
and a non-equilibrated  ($\lambda_q$=0.2 and 0.02) plasma.
As can be seen, the rate of dilepton production is not
sensitive to $x$ when the plasma is highly unsaturated (the
lower curve for $\lambda_q=0.02$). However, in case of an equilibrated plasma,
it starts deviating more and more for higher $x$ values (upper curve for $\lambda_q$=1.0).
Since the plasma is highly unsaturated at the beginning ($\lambda_q \approx .02$)
and reaches up to $\lambda_q \approx .2$ as it equilibrates, the rate of
dilepton production will be insensitive to the small values of baryo-chemical
potential.
It can be mentioned here that  we have also estimated the thermal dilepton
production rate using the approximation (11). We do not find much difference
with the actual calculations particularly when the plasma is unsaturated.
This further justifies the use of the MFD approximation both for quark and anti-quark
distributions.
Next we calculate the integral over the space and time
using the expression:

\equation \frac {dN}{dydM^2} = \pi R_A^2
\int_{\tau_0}^{\tau_c} d \tau \tau ~\frac{dR}{dM^2}\endequation
The rate dR/dM$^2$ is estimated numerically using the exact distribution
functions.
However, for the space time evolution of the plasma at finite baryon density
(i.e. for $T(\tau)$
and $\lambda_q(\tau)$), we use the results of the previous section
which is obtained using the MFD approximation.
Fig. 5(b) shows the  above integrated yields
at a fixed initial temperature of $T_0=0.57$ GeV and initial fugacities  $\lambda_{g0}=0.09$,
$\lambda_{q0}=\lambda_{{\bar q}0}=0.02$ for different values of $x_0$.
Although, the rate of dilepton production do not depend on $x$ strongly,
the space time integrated yield
increases with $x_0$
due to slower cooling of the plasma in the presence of baryon density .
Since the energy density is directly related to the
experimental observables, we have
also calculated the dilepton yield at a fixed initial energy density of
$\epsilon_0=9$ GeV/fm$^3$.
This energy density corresponds to an initial temperature of $T_0=0.57$ GeV
at $\lambda_{g0}$=0.09 and
$\lambda_{q0}=0.02$ when $x_0=0$. The results are shown in Figs. 6(a) and 6(b)
for both the cases of equilibrated
and non-equilibrated plasma.
As shown in the figures, if the plasma is chemically non-equilibrated,
the suppression
of the dilepton yield at higher baryo-chemical potential
is not significant even at higher invariant masses in contrast to the
case for a chemically equilibrated plasma.

\section {Conclusion}
In the present work, we have studied the effect of finite baryon density
on the chemical equilibration of a
longitudinally expanding quark gluon plasma.
It is found that the rate of chemical equilibration for gluon
slows down in the presence of finite baryon density
in comparison to a baryon-free plasma,
irrespective of various assumptions about the equilibration mechanism .
This results in a slower cooling of the plasma , which has an important consequence
resulting in  higher dilepton yields
even though the rate of production may be suppressed at finite baryon density
due to lower initial temperature. We have
studied the thermal dilepton yields from the chemically non-equilibrated
expanding plasma with finite baryon density .
For calculating  the dilepton yields,
we have considered only an ideal non-viscous fluid
that undergoes isentropic expansion and  we have neglected higher order gluon
processes as well as  temperature dependence of the coupling constant.
It is found that the space-time integrated yields of dileptons
are enhanced if the initial temperature is held fixed. More importantly,
even for a fixed
initial energy density, the suppression of the yield at higher baryo-chemical
potential is compensated to a large extent by the slow cooling of the plasma.

\section*{Acknowledgement} We are thankful to Dr. D. K. Srivastava for illuminating
discussions on several aspects of the present study.
\setcounter{equation}{0}
\section*{Appendix A}
Figs. A.1 and A.2 show the quark distribution function for p=0 and p=0.5 GeV
as a function of $\lambda_q$ at $x$=1.0.
Figs. A.3 and A.4
show the corresponding plots for the anti-quark distributions. In all the
cases, the FD approximations
deviate significantly from the true
Juttner distribution except when $\lambda \rightarrow 1$.
This deviation is more significant in case of the baryon free plasma
although the FD approximation has been used widely in all earlier calculations.
At $x$=0, the BM distribution
(not shown in the figure), is found to be
closer to the Juttner distribution at small fugacities, but it starts
deviating even at small $\lambda$ when baryon density increases. On the other
hand, the MFD approximation is relatively closer to the actual distribution
over a wide range of fugacities (A.1 and A.2). However, the anti-quark
distribution function (A.3 and A.4) becomes closer to BM approximation
as $x$ increases.
At any momentum, for an unsaturated plasma at finite baryon density,
the quark distribution function
can be best described by the MFD approximation whereas the BM
approximation is more
suitable for the anti-quark distribution function. However, we do not want to use
the MFD for $f_q$ and BM for $f_{\bar q}$ as they will result in  baryon
asymmetry when $x \rightarrow 0$. Moreover, at finite baryon density, it is
the $f_q$ distribution which dominates as the anti-quarks are strongly
suppressed (A.4). Therefore, we use the MFD approximation both for
$f_q$ and $f_{\bar q}$ in order to evaluate $m_q^2$, $m_D^2$ and quark and anti-quark
energy and number densities. We have also solved the rate equations
using BM approximation both for quark and anti-quark distributions,
but the final results do not change much . This is primarily due to the fact
that the $x$ dependence comes through the exponential $e^{\pm x}$ in the
numerator of eq. (1) and it does not matter if one uses BM or MFD approximations.
However, we have used the MFD approximation for the calculations so that
the same parametrization
of ref. \cite{5} can be employed by solving the rate equations.
\setcounter{equation}{0}
\section *{Appendix B}
\renewcommand{\theequation}{B.\arabic{equation}}
    For a baryon free plasma ($\mu=0$), Biro et. al. have used a simple
 factorisation for the rate equations. We can retain the same factorisation
 for both the processes $gg\rightleftharpoons q\bar{q}$ and
$gg\rightleftharpoons ggg$ (eq.(3) and (4)) under the MFD approximations.
 Using the identity (7,8) and the MFD approximations (11,12) for quark, anti-quark
 and gluons, the gain and
loss terms for the process $gg\rightleftharpoons ggg$ (eq. (5) and (6))
can be combined to give
\begin{eqnarray}
R_{g \rightarrow g}-R_{q \rightarrow g} =&(\lambda_g^2-\lambda_Q \lambda_{\bar Q})
\frac{1}{2} \int \frac{d^3p_1}{(2\pi)^3 2 E_1}
\int \frac{d^3p_2}{(2\pi)^3 2 E_2} \int \frac{d^3p_3}{(2\pi)^3 2 E_3}
\int \frac{d^3p_4}{(2\pi)^3 2 E_4}~~~~~\times \nonumber\\
&(2\pi)^4 \delta^4(p_1+p_2-p_3-p_4)
\sum |M|^2  f_q^{eq}(p_3) f_{\bar q}^{eq}(p_4) f_g^{eq}(p_1) f_g^{eq}(p_2)
e^{\frac {E_1+E_2}{T}}
\end{eqnarray}
where $p_1+p_2 = E_1+E_2$ for mass less quarks or gluons.
From the above expression, it is clear that the right hand side vanishes
when $\lambda_g$= $\lambda_q$= $\lambda_{\bar q}$=1.
Since the plasma is highly unsaturated, it may be reasonable to replace
$f^{eq}$ with the Boltzmann distribution. The integral in
eq. (B.1) can be written as
\begin{equation}
I = \frac{1}{2} \int \frac{d^3p_1}{(2\pi)^3 }
\int \frac{d^3p_2}{(2\pi)^3 }
[\sigma_{gg \rightarrow q\bar q} v_{12}] f_g(p_1) f_g(p_2)
\end{equation}
which represents the free space cross section for the process $ gg \rightarrow q\bar q$
folded with the distributions for the initial particles. The cross section
$\sigma_{gg \rightarrow q\bar q}$ is given by
\equation
\sigma_{gg \rightarrow q\bar q}= \frac{1}{v_{12} 2E_1 2E_2}
\int \frac{d^3p_3}{(2\pi)^3 2 E_3} \int \frac{d^3p_4}{(2\pi)^3 2 E_4}
(2\pi)^4 \delta^4(p_1+p_2-p_3-p_4) \sum |M|^2
\endequation
  It can be mentioned here that the above integral (B.2)
  is identical to the one which could have been obtained
with the classical approximations, i.e. using the Boltzmann distribution function
for quark, anti-quark and gluon and eliminating the Pauli blocking and
Bose enhancement factors in the final states in (5) and (6). Although, we use
the quantum statistics, the same expression is obtained as the classical one due to
the identity and the approximations eqs. (7,8,11,12). However Boltzmann approximation
is used finally for $f^{eq}$.

Following \cite{5}, eq. (B.2) can be factorised as
\equation
R_{g \rightarrow g} - R_{q \rightarrow g}=\frac{1}{2}\sigma_2~n_g^2(1-\frac
{\lambda_Q\lambda_{\bar Q}}{\lambda_g^2})
\endequation
Similarly for the rate $gg \rightarrow ggg$, one can use
\begin{equation}
R_{2 \rightarrow 3} - R_{3 \rightarrow 2}=\frac{1}{2}\sigma_3~n_g^2(1- \lambda_g)
\end{equation}
where
$$\sigma_2=<\sigma(gg\rightarrow q\bar q)v>,~~\sigma_3=<\sigma(gg\rightarrow q\bar q)v>$$

\section{Appendix C}
\setcounter{equation}{0}
The quark gluon plasma in (1+1) dimenison expands following the scaling
law $T^3\tau=const$. If, chemical equilibration is not complete,
the plasma cools faster than $t^{-1/3}$ as additional energy is spent in
chemical equilibration. Therefore, in case of a chemically equilibrating
plasma, the cooling rate depends on the rate of equilibration of quarks and
gluons present in the plasma. Further, as shown in the text,
the presence of baryo-chemical potential makes the gluon equilbration slow
and also the temperature of the plasma falls slowly in comparison to the
baryon free case. However, as will be shown here, the slow cooling of the
plasma is not entirely due to the decrease of the gluon equilibration rate, but
also due to the presence of baryo-chemical potential which effects the hydrodynamical
expansion of the plasma. This has been demonstrated in fig. C where the temperature
T, the gluon ($\lambda_g$) and quark ($\lambda_q$) fugacities have been plotted
as a function of $\tau$ both for baryon free and baryon rich cases ($x$=0 and
2.0, see the solid and dashed curves). As discussed in the text, the gluon
equilibration rate slows down and the plasma cools slowly when the baryon
density is finite where as the quark equilibration rate is not effected much.
Next, we consider only a baryon free plasma, but we try to reduce the gluon and
the quark equilibration rate by reducing the rate $R_3$ and $R_2$ by a factor
f=0.6 so that the plasma now cools with the same rate (see the dotted line)
as that of baryon rich case
with $x$=2.0 (dashed curve obtained using the normal value of $R_2$ and $R_3$).
However, the equilibration rates for gluon and quarks now become much slower (dotted
lines) as compared to the dashed curves. This indicates that if the plasma
needs to cool with a rate as that of dashed curve just due to chemical
equilibration alone
, the gluon and quarks need to equilibrate with a rate much slower than what is
shown in fig. C corresponding to the case of a baryon rich plasma.
In other words, a higher equilibration rate of gluon and quark as that of dashed
curves will consume more energy and the plasma will cool faster than what is
shown by dashed curve (but still slower than the solid line) had the
chemical equilibration been the only reason for
the deviation of the cooling of the plasma from the Bjorken's scaling.
Therefore, in  presence of baryon density, the slow cooling rate of the plasma
arises due to two factors; one being the
slow gluon equilibration rate which consumes less energy. The other one
is due to the slow hydro-dynamical expansion as the plama now needs to expand
conserving the baryon number as well.

\newpage
\begin{itemize}
\item Fig. 1 \\
      (a) The product of quark and anti-quark distribution function as a function
      of p for different values of $x$ at $T=0.57$ GeV and 0.2 GeV.
      (b) Same as (a), but as a function of $\lambda_q$.

\item Fig. 2\\
      The thermal quark  mass $m_q^2/T^2$ versus
      $\lambda_q$ at $x=1.0$ and $x=1.5$
      (the upper curves are increased by one unit).
      The solid curve is for the Juttner distribution function
      (eq. 1 ). The other curves are obtained using different
      approximations, MFD (eq. 11), FD (eq. 13) and BM (eq. 14) respectively.

\item Fig. 3\\
      The gluon production rate, $R_3/T$, and quark production rate $R_2/T$
      as function of $\lambda_g$ $(\lambda_q =\lambda_{\bar q}= \lambda_g/5)$
      for  $x=0.0$, 1.0 and 1.5 . The dashed dot curve is for $R_3/T$
      at $x=0.0$ with $m_D^2= 4 \pi \alpha_s ~T^2~\lambda_g$
      as used in ref. \cite{7}.

\item Fig. 4\\
      (a) The temperature, (b) the quark chemical potential $x=\mu /T$,
      (c) the gluon fugacity $\lambda_g$, (d) the quark fugacity $\lambda_q$,
      as a function of $\tau$ for $x_0=0.0$, 1.0 and 1.5 with
      the initial conditions  $T_0=0.57$ GeV, $\lambda_{g0}=0.09$ and
      $\lambda_{q0}=0.02$. The dotted line in (a)
      corresponds to the temperature as per the Bjorken's scaling.
\item Fig. 5\\
       (a) The dilepton production rate $dR/dM^2$ as a function of invariant
       mass M for  $\lambda_q= 0.02$, 0.2 and 1.0 . The initial temperature
       of the plasma is $T_0= 0.57$ GeV.
       (b) The space time integrated dilepton yield versus invariant mass M
       for a
       chemically non-equilibrated plasma ($\lambda_{g0}$=0.09, $\lambda_{q0}$=0.02 )
        at a fixed initial temperature of $T_0=0.57$ GeV with $x_0=0.0$, 1.0 and 1.5.
\item Fig. 6\\
        (a) The integrated dilepton yield from a chemically equilibrated plasma
        at a fixed initial energy density for various values of $x_0$.
        (b) Same as Fig. 6(a),  but from a chemically non-equilibrated plasma.

\item Fig. A\\
       The quark ($f_q$) and anti-quark ($f_{\bar q}$)
       distribution functions versus
       $\lambda_q$ at $p=0$ (A.1 and A.3) and $p=0.5$ GeV (A.2 and A.4).
\end{itemize}

\newpage
\begin{figure}[h!]
\vspace{5cm}
\hspace{3cm}
\begin{minipage}[t]{12.0cm}
\psfig{figure=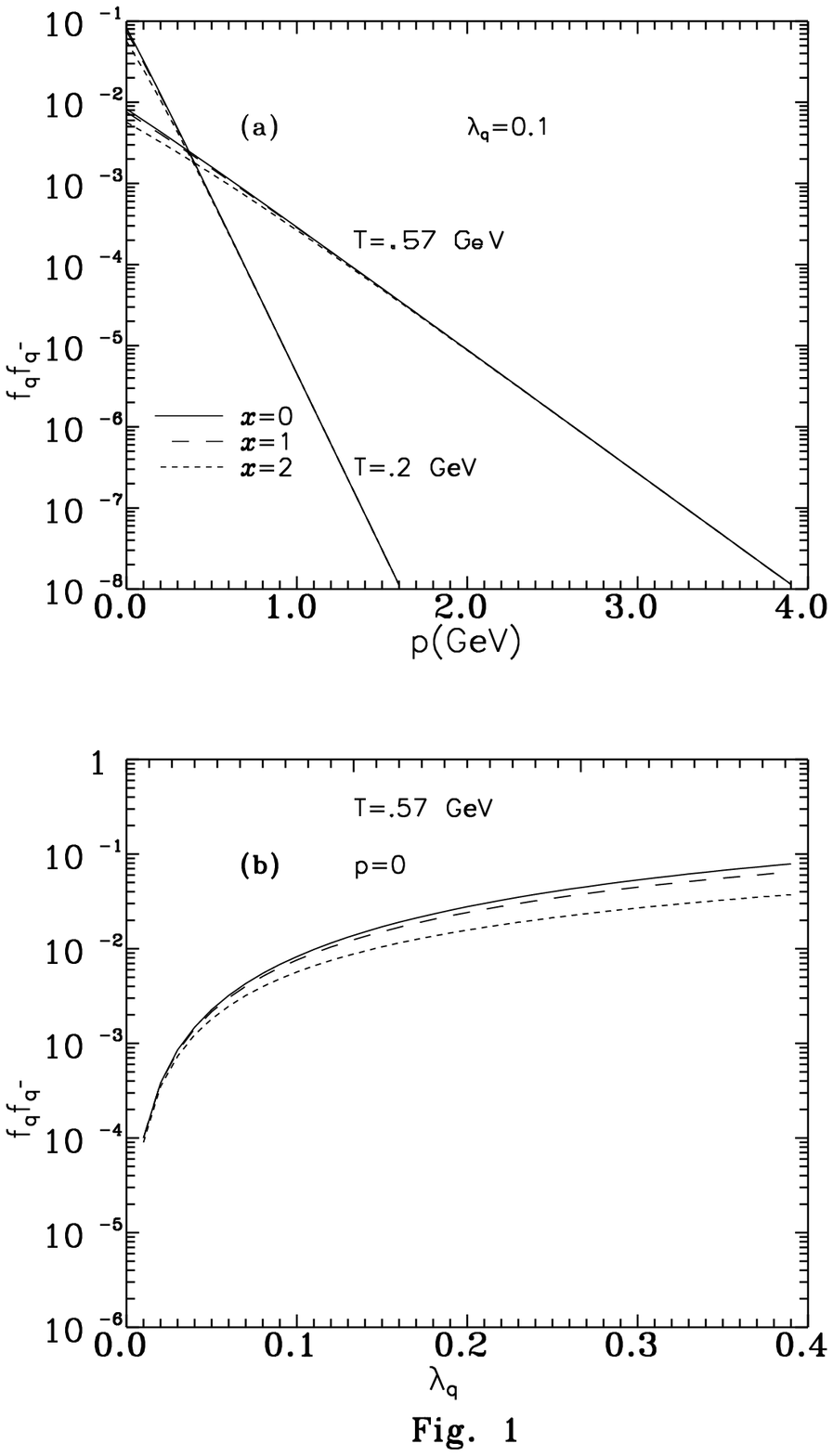,width=12.0cm,height=12.0cm}
\end{minipage}
\end{figure}
\newpage
\begin{figure}[h!]
\begin{minipage}[t]{12.0cm}
\psfig{figure=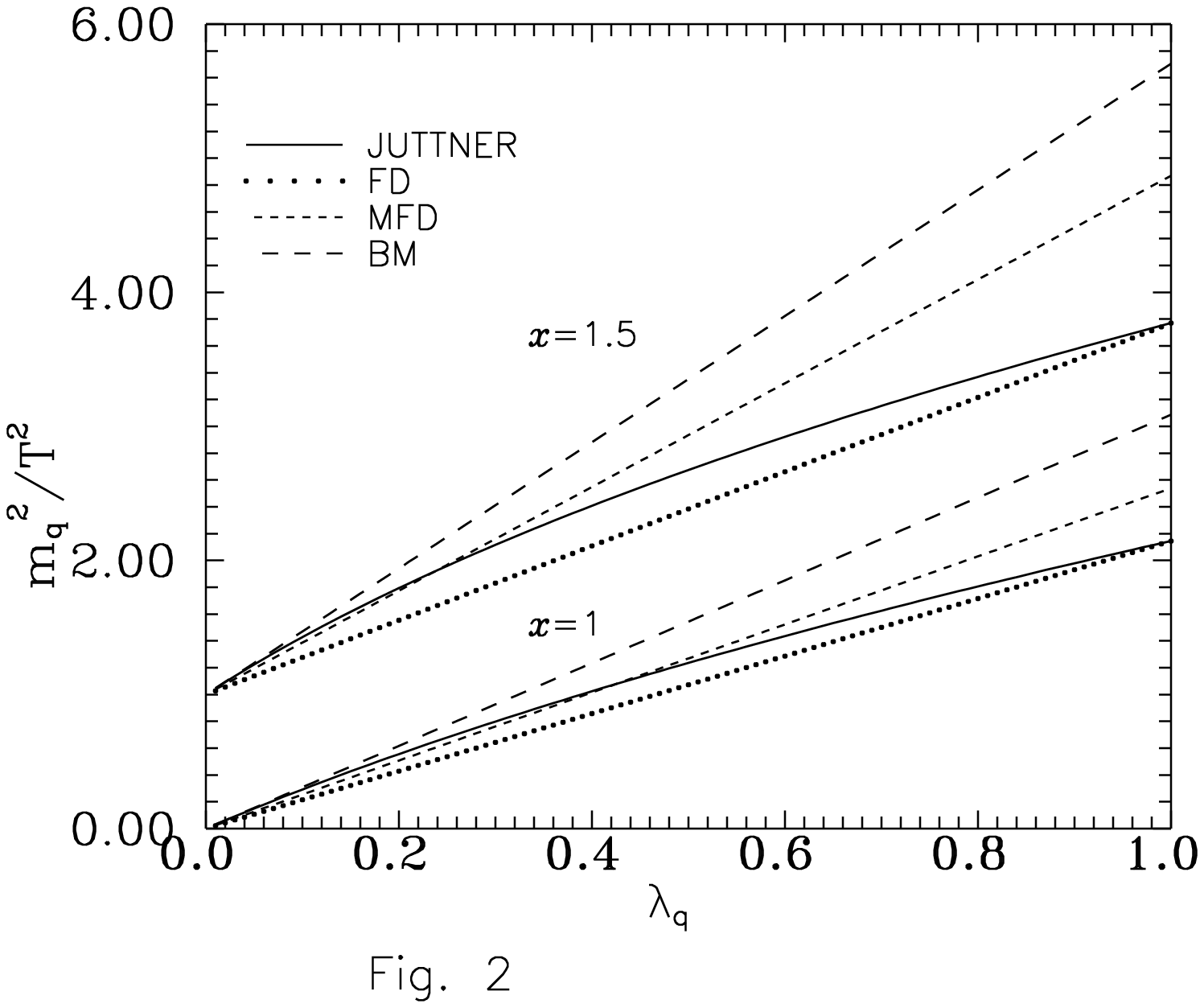,width=12.0cm,height=12.0cm}
\end{minipage}
\end{figure}
\newpage
\begin{figure}[h!]
\vspace{3cm}
\begin{minipage}[t]{12.0cm}
\psfig{figure=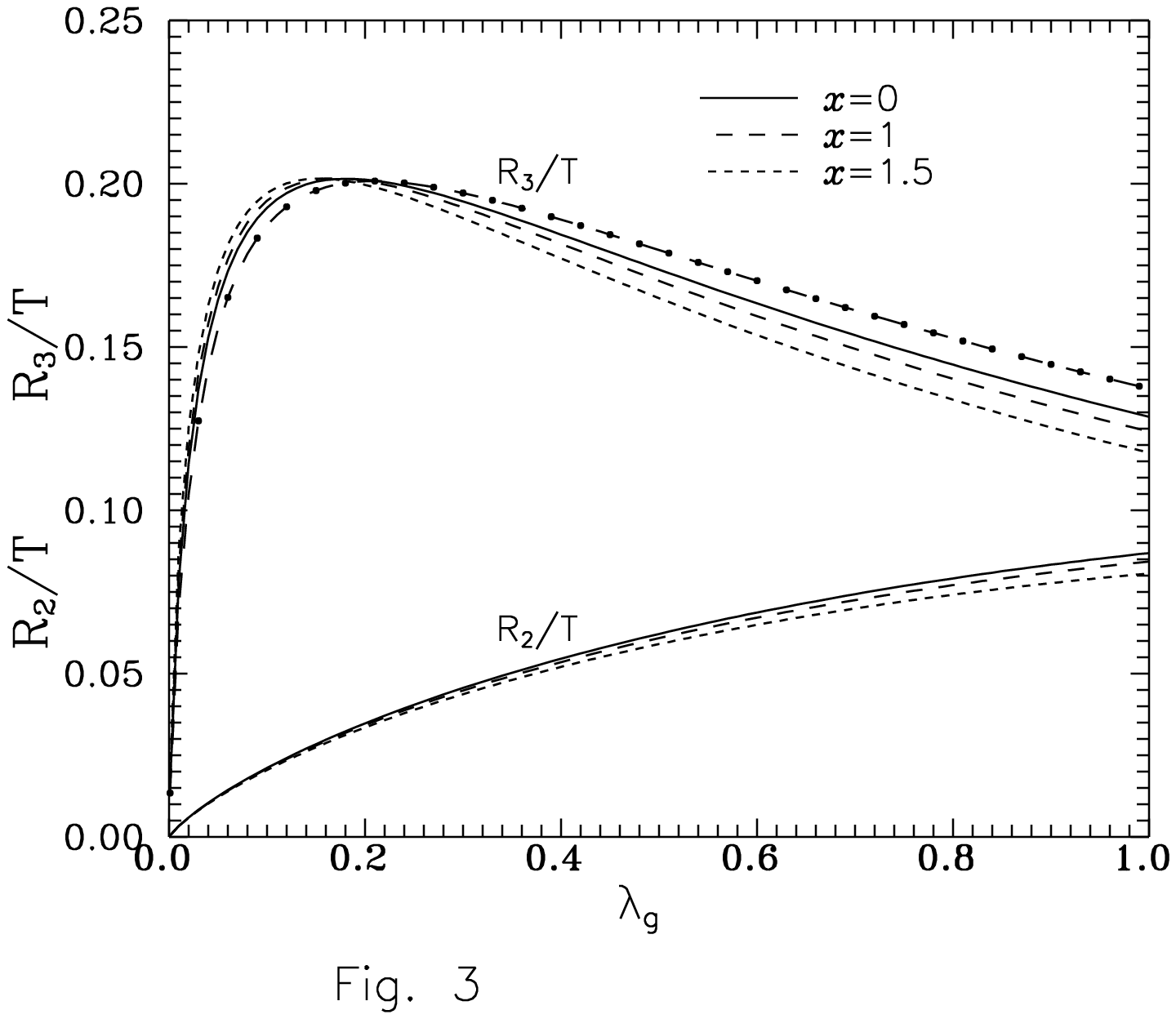,width=12.0cm,height=12.0cm}
\end{minipage}
\end{figure}
\newpage
\begin{figure}[t!]
\vspace{3cm}
\begin{minipage}[t]{12.0cm}
\psfig{figure=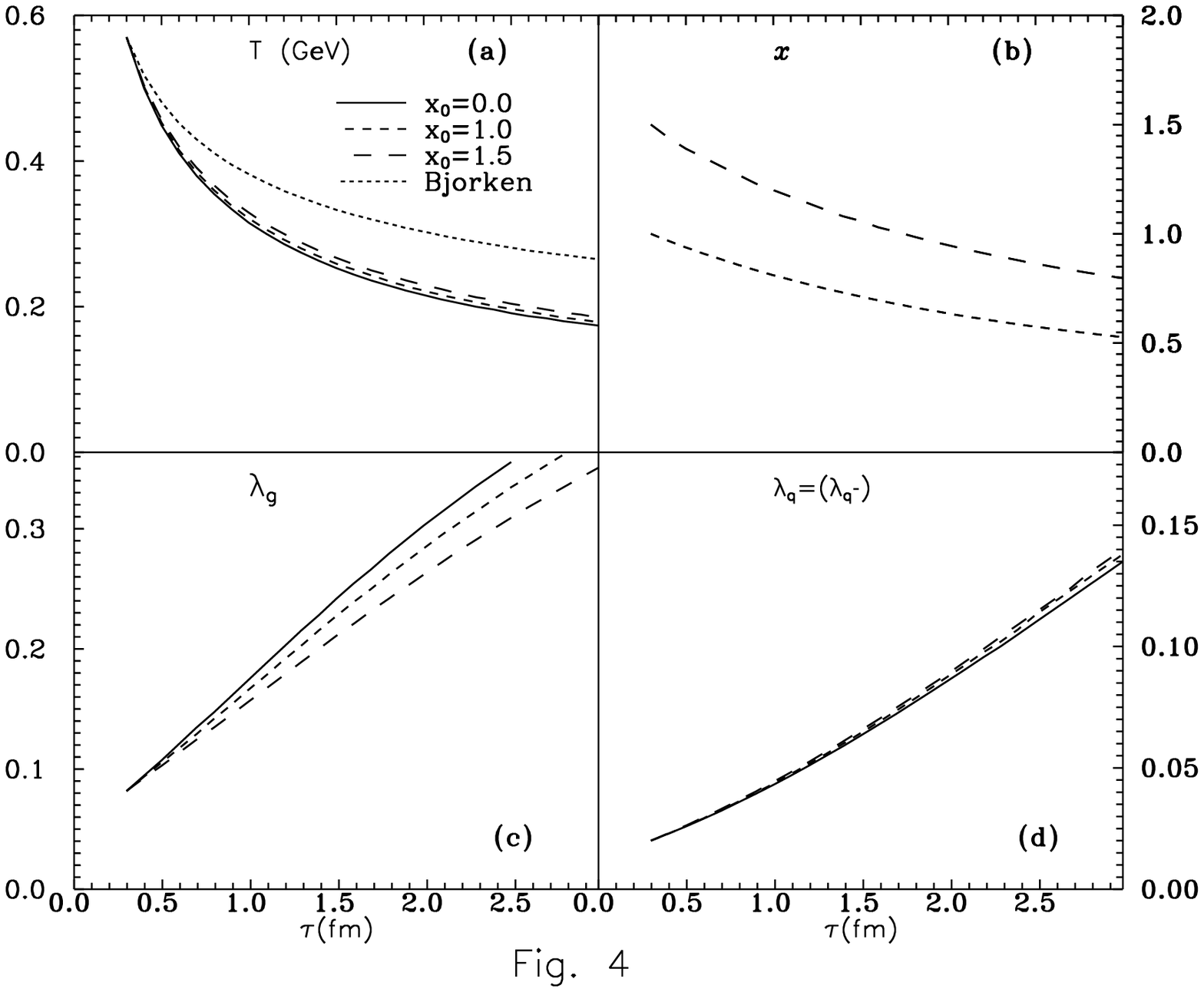,width=12.0cm,height=12.0cm}
\end{minipage}
\end{figure}
\newpage
\begin{figure}[t!]
\vspace{3cm}
\begin{minipage}[t]{12.0cm}
\psfig{figure=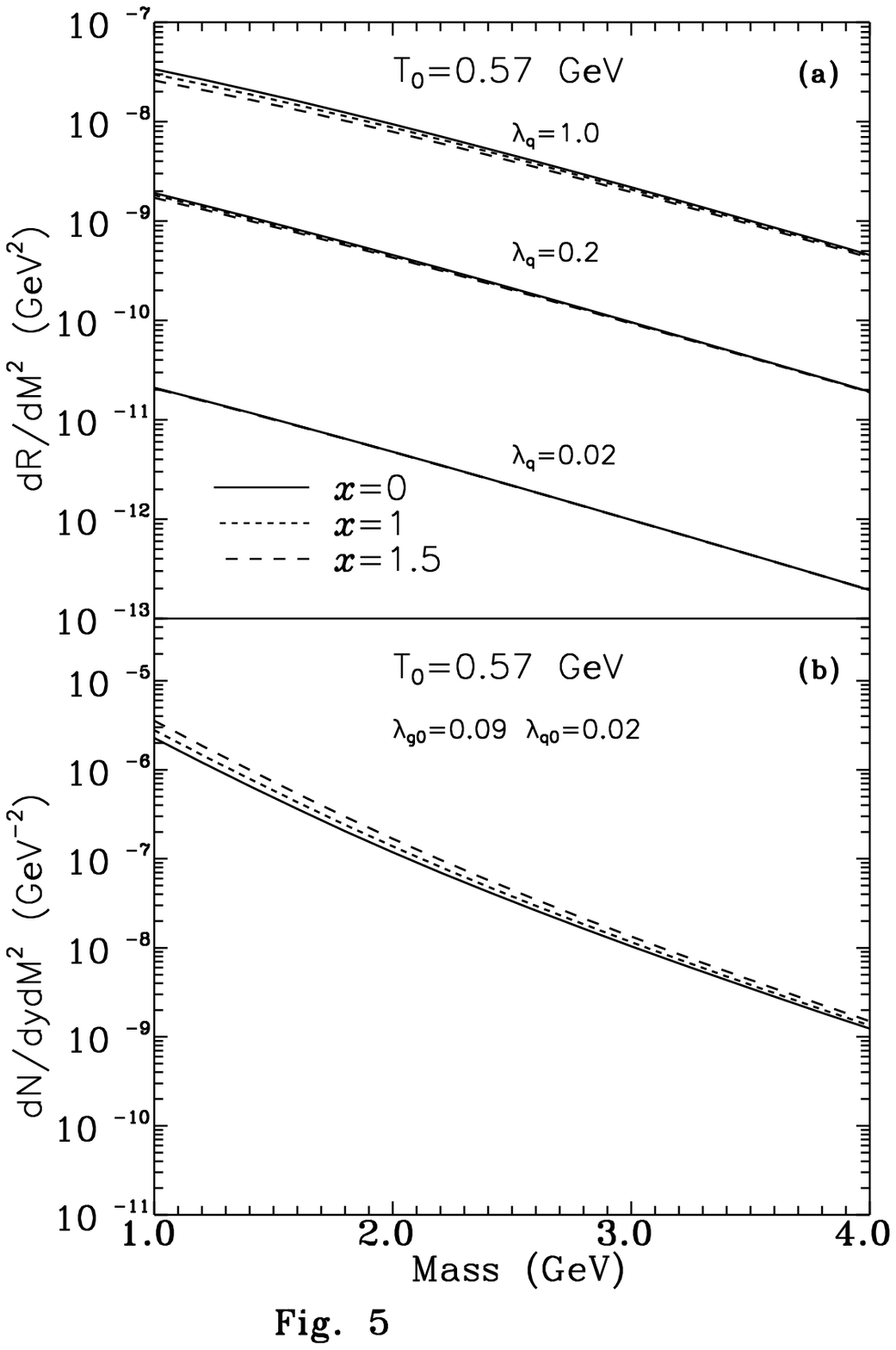,width=12.0cm,height=12.0cm}
\end{minipage}
\end{figure}
\newpage
\begin{figure}[t!]
\vspace{3cm}
\begin{minipage}[t]{12.0cm}
\psfig{figure=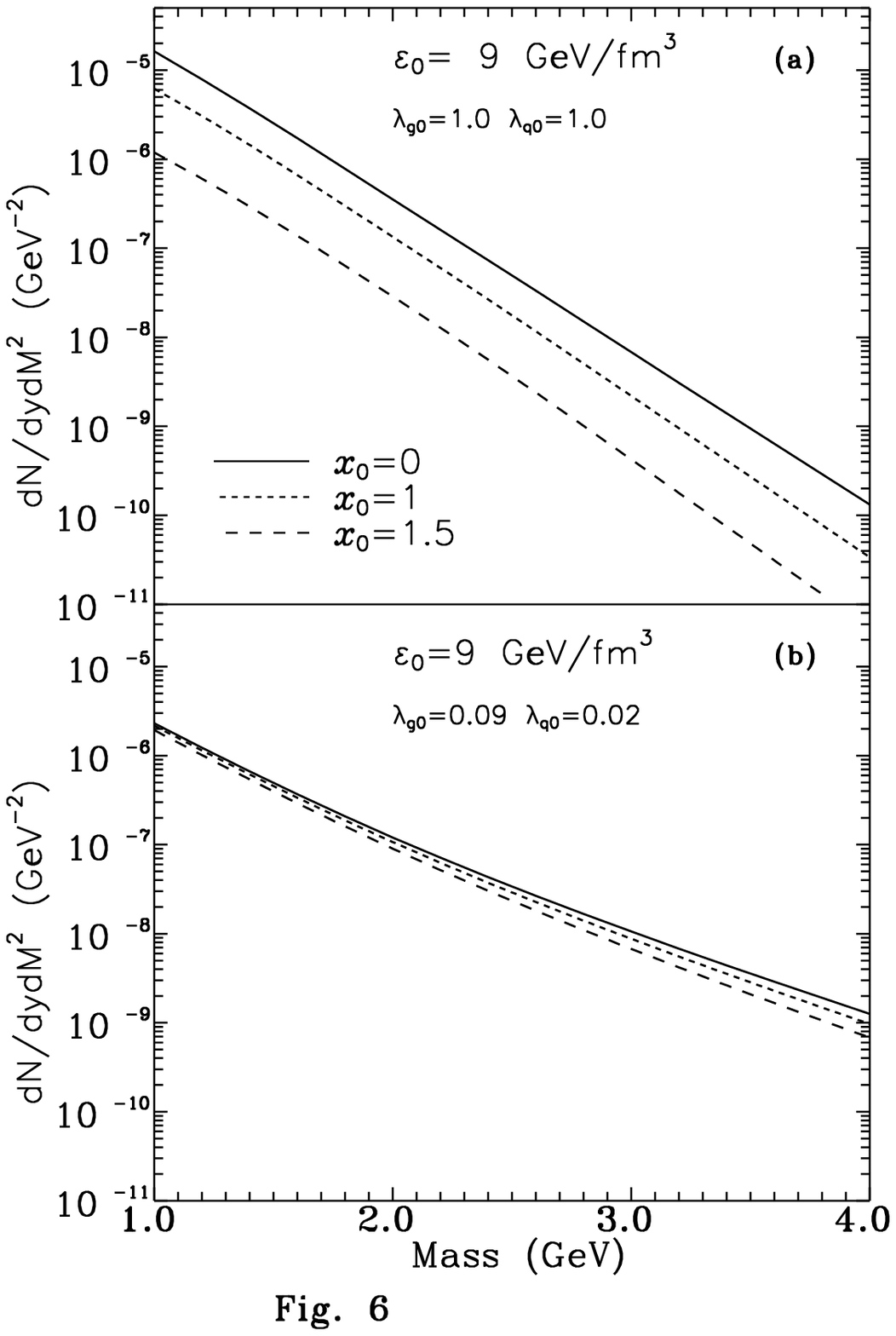,width=12.0cm,height=12.0cm}
\end{minipage}
\end{figure}
\newpage
\begin{figure}[t!]
\begin{minipage}[t]{12.0cm}
\psfig{figure=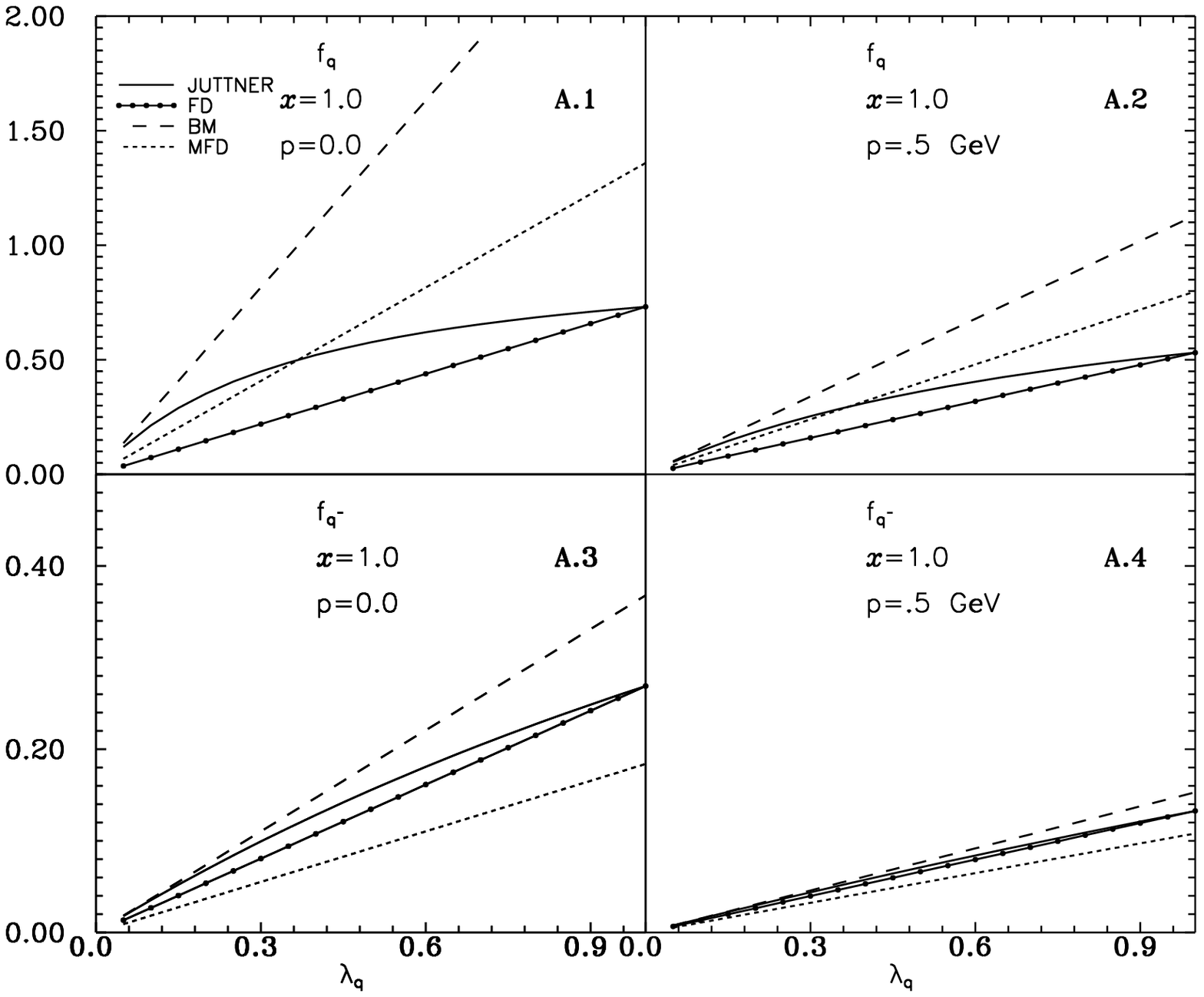,width=12.0cm,height=12.0cm}
\end{minipage}
\end{figure}
\newpage
\begin{figure}[t!]
\begin{minipage}[t]{12.0cm}
\psfig{figure=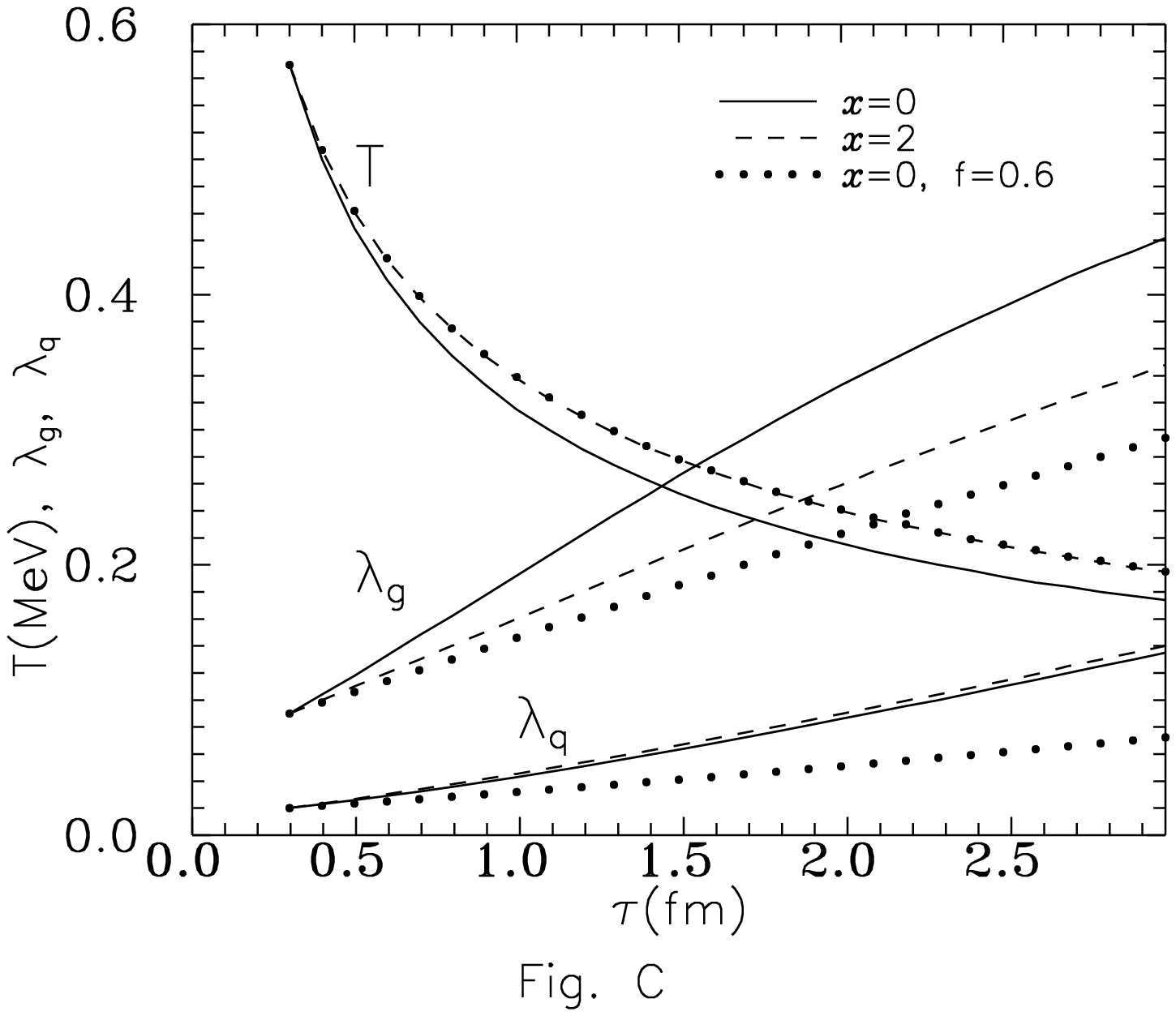,width=12.0cm,height=12.0cm}
\end{minipage}
\end{figure}

\end{document}